\documentclass{article}
\usepackage{arxiv}
\makeatletter
\expandafter\let\csname table*\endcsname\table
\expandafter\let\csname endtable*\endcsname\endtable
\makeatother
\usepackage{cite}

\usepackage{amsmath,amssymb,amsfonts}
\usepackage{algorithmic}
\usepackage{graphicx}
\usepackage{textcomp}
\usepackage{xcolor}
\definecolor{rlblue}{HTML}{2A78D6}
\definecolor{llmorange}{HTML}{E2662F}
\usepackage{booktabs}
\usepackage{tikz}
\usetikzlibrary{arrows.meta,positioning,shapes.geometric,fit,backgrounds}
\graphicspath{{figs/}}
\def\BibTeX{{\rm B\kern-.05em{\sc i\kern-.025em b}\kern-.08em
    T\kern-.1667em\lower.7ex\hbox{E}\kern-.125emX}}
\renewcommand{\shorttitle}{No One Architecture Fits All}
\date{}
\begin{document}

\title{No One Architecture Fits All: A Cross-Environment Evaluation of Hierarchical Red Team Agents
}

\author{Ayan Javeed Shaikh \\
\textit{Indiana University} \\
Bloomington, IN, USA \\
ayshaikh@iu.edu
\And
Arunesh Sinha \\
\textit{Rutgers University} \\
Piscataway, NJ, USA \\
arunesh.sinha@rutgers.edu
\And
Nathaniel D. Bastian \\
\textit{Johns Hopkins University} \\
Baltimore, MD, USA \\
ndbastian@jhu.edu
\And
Ankit Shah \footnote{}\\
\textit{Indiana University} \\
Bloomington, IN, USA \\
ankit@iu.edu
}
\maketitle
{\renewcommand{\thefootnote}{}\footnotetext{Send correspondence to ankit@iu.edu \\ \copyright~2026 IEEE. Personal use of this material is permitted. Permission from IEEE must be obtained for all other uses, in any current or future media, including reprinting/republishing this material for advertising or promotional purposes, creating new collective works, for resale or redistribution to servers or lists, or reuse of any copyrighted component of this work in other works.}}

\begin{abstract}
Autonomous red team agents increasingly stress-test AI-enabled cyber defenses by planning strategy and executing multistage attacks. Reinforcement learning (RL) and large language models (LLMs) offer complementary mechanisms for the planning and execution such agents require, and prior work has combined them in hybrid hierarchies. Yet a given architecture is typically developed and evaluated within a single environment, leaving open whether an observed advantage reflects a generally stronger decision mechanism or merely alignment with a particular setting. We address this gap with a controlled cross-environment comparison of two homogeneous hierarchical red team architectures: an RL planner with an RL executor (RL+RL) and an LLM planner with an LLM executor (LLM+LLM). We evaluate both against expert autonomous defenders in CybORG CAGE-4 and in Cyberwheel at two network scales, across 18 configurations under one unified disruption metric. We find a pronounced environment-dependent inversion. {\color{black}RL+RL wins the compact, densely rewarded CAGE-4 (78.5\% disruption success versus 18.0\% for the strongest LLM configuration) and the 100-host Cyberwheel network (81.0\% versus 50.5\%), while a pretrained cybersecurity LLM agent wins the larger, escalation-gated 1010-host Cyberwheel network (55.0\% versus 0.0\% for RL).} A kill-chain analysis explains the inversion through architecture-specific bottlenecks that aggregate success rates conceal. {\color{black}In the 1010-host Cyberwheel network, RL discovers and compromises hosts but stalls at privilege escalation, whereas in CAGE-4, LLM agents obtain privileged access but rarely convert it into operational impact.} These results indicate that conclusions drawn in a single environment may not generalize, and that hybrid planner-executor designs should be motivated by specific failure modes rather than the assumption that one architecture is universally preferable.
\end{abstract}


\section{Introduction}
Autonomous cyber operations are moving beyond fixed scripts toward agents capable of interpreting network state, selecting objectives, and executing multistage actions. Recent advances in reinforcement learning (RL), multi-agent learning, and large language models (LLMs) have accelerated this transition \cite{kiely2025marl,singh2025hierarchical,jiang2025agentic,deng2024pentestgpt,gioacchini2025autopenbench}. These capabilities are particularly valuable for automated red teaming, where autonomous adversaries can systematically stress-test cyber defenses before weaknesses are encountered operationally. Effective red teaming, however, requires more than discovering vulnerable hosts or executing isolated exploits; an agent must maintain strategic direction while progressing through reconnaissance, exploitation, privilege escalation, lateral movement, and operational impact.

RL and LLMs offer fundamentally different mechanisms for supporting such behavior. RL agents learn sequential policies through repeated environment interaction and can acquire efficient, specialized tactical behaviors, while hierarchical and multi-agent RL decompose complex cyber operations into coordinated high- and low-level decisions \cite{kiely2025marl,singh2025hierarchical,jiang2025agentic}. LLM agents instead leverage pretrained knowledge and language-based reasoning to interpret observations, formulate plans, and select actions without learning every behavior directly from interaction. Recent work has demonstrated their potential for penetration testing, vulnerability exploitation, and multistage cyber reasoning \cite{deng2024pentestgpt,gioacchini2025autopenbench,zhu2026zeroday}. However, RL learned policies may transfer poorly when operational dynamics differ from training, whereas LLM agents can struggle with long-horizon context, executable action selection, and recovery from failed decisions \cite{deng2024pentestgpt,gioacchini2025autopenbench,zhu2026zeroday}.

The complementary strengths of these paradigms have also motivated hybrid approaches that combine LLM-based strategic reasoning with RL-based tactical execution or use LLM guidance within hierarchical RL architectures \cite{jiang2025agentic,shaikh2026spie}. While such studies show that the two paradigms can complement one another, they leave a more fundamental question unresolved: \emph{how do hierarchical systems built around learned policies versus language model reasoning compare across different cyber environments?} Existing peer-reviewed studies typically develop and evaluate a particular RL, LLM, or hybrid architecture within a specific environment or task. Hence, limited evidence exists on whether an observed architectural advantage reflects a generally stronger decision mechanism or simply better alignment with a particular environment.

This distinction matters because cyber environments can differ substantially in topology, scale, action semantics, privilege requirements, and defensive behavior. Additionally, aggregate measures such as reward or overall attack success can conceal different failure mechanisms. An agent may discover and compromise hosts but fail at privilege escalation, while another may obtain privileged access yet fail to convert it into operational impact. Although recent LLM-agent benchmarks have begun to examine intermediate milestones \cite{gioacchini2025autopenbench}, controlled cross-environment comparisons of hierarchical RL-based and LLM-based red agents remain limited.

To address this gap, we compare two homogeneous hierarchical red agent architectures: an RL planner coupled with an RL executor (RL+RL) and an LLM planner coupled with an LLM executor (LLM+LLM). We treat these configurations as architectural endpoints to characterize the strengths and failure modes of learned-policy and language-model-based decision making. Both approaches are evaluated against expert autonomous defenders in CybORG CAGE-4~\cite{kiely2025marl}, as well as on Cyberwheel~\cite{cyberwheel} networks at mid-scale with 100 hosts and large-scale with 1010 hosts. Across 18 evaluation configurations and 3{,}600 attack episodes, we assess disruption success, attack chain progression, behavioral efficiency, and planning cost.

Our results reveal a pronounced environment-dependent inversion in architectural effectiveness. In CAGE-4, RL+RL reaches a disruption success rate of 78.5\%, compared with 18.0\% for the strongest LLM configuration. {\color{black}In Cyberwheel, the outcome differs across the two network scales: RL+RL also leads in the 100-host environment (81.0\% versus 50.5\% for LLM+LLM), whereas the relationship reverses in the 1010-host environment, where LLM+LLM achieves 55.0\% and RL+RL achieves 0.0\%.} Kill-chain analysis further exposes distinct downstream bottlenecks. {\color{black}RL agents in the 1010-host Cyberwheel environment discover and compromise hosts but largely stall at privilege escalation, whereas LLM agents in CAGE-4 frequently obtain privileged access but struggle to convert it into operational impact.} These findings show that conclusions drawn from one cyber environment or network size may not generalize to substantially different settings.

This work makes four contributions:
\begin{itemize}
\item \textbf{Hierarchical cross-paradigm comparison:} a controlled comparison of RL+RL and LLM+LLM red agent architectures as distinct end-to-end decision paradigms.
\item \textbf{Cross-environment evaluation:} evaluation across CAGE-4 and two Cyberwheel scales, revealing that relative architectural effectiveness reverses across settings.
\item \textbf{Failure-mode characterization:} analysis of attack progression from discovery through privilege escalation and disruption, identifying architecture-specific bottlenecks hidden by aggregate success measures.
\item \textbf{Implications for hybrid design:} empirical evidence that future LLM-RL hybrids should be motivated by specific planning and execution failure modes rather than an assumption that one architecture is universally preferable.
\end{itemize}

The rest of the paper is organized as follows. We first review related work, followed by describing the architectures and environments, then presenting the experimental results and analysis, and finally, concluding with future directions.

\section{Background and Related Work}
\textbf{Autonomous cyber environments and defenders.} Simulation gyms are the standard proving ground for autonomous cyber agents. CybORG models enterprise networks under partial observability \cite{baillie2020cyborg}, and its CAGE Challenge 4 (CAGE-4) defines a multi-subnet, multi-agent scenario with coordinated defenders \cite{kiely2025marl}. The strongest published CAGE-4 defender is a Hierarchical Multi-Agent RL (H-MARL) expert policy \cite{singh2025hierarchical}, and graph-based RL agents target the same benchmark \cite{king2025automated}. Cyberwheel is a complementary RL simulation environment for autonomous cyber defense that models configurable networks at substantially larger scale \cite{cyberwheel}. Red and blue agents are almost always developed and evaluated within a single environment, so it is rarely possible to tell whether a reported advantage transfers to a structurally different network.

\textbf{Language models for offensive cyber.} LLMs have advanced quickly on offensive tasks. PentestGPT decomposes penetration testing into reasoning, generation, and parsing modules \cite{deng2024pentestgpt}; benchmarks such as AutoPenBench evaluate generative agents on staged tasks and surface their intermediate milestones \cite{gioacchini2025autopenbench}; and teams of LLM agents can exploit zero-day vulnerabilities \cite{zhu2026zeroday}. HackSynth pairs a planner and summarizer for autonomous penetration testing \cite{muzsai2024hacksynth}, and prompted LLM agents can rival RL agents in network security games \cite{rigaki2024stochastic}. Domain-specialized models such as Foundation-Sec-8B \cite{foundation_sec} and DeepHat, formerly WhiteRabbitNeo \cite{deephat}, sharpen security reasoning through targeted fine-tuning. Even so, standalone LLMs struggle to sustain long-horizon campaigns under active defense; evaluated as autonomous agents in CAGE-4, all-LLM teams trail learned policies substantially \cite{castro2025llm_acd}.

\textbf{Reinforcement learning and hierarchy for cyber.} RL learns attack and defense policies directly from interaction, commonly with policy gradient methods such as proximal policy optimization (PPO) \cite{schulman2017ppo}. Multi-agent and hierarchical RL decompose enterprise-scale operations into coordinated high- and low-level decisions, which is effective for CAGE-4 defense \cite{kiely2025marl,singh2025hierarchical}. Such policies excel at precise tactical execution but are sample-inefficient and transfer poorly when the network or defender differs from training.

\textbf{Hybrid architectures and the gap we address.} A growing line of work combines the two paradigms: LLM-guided hierarchical MARL directs a learned hierarchy with a language model \cite{jiang2025agentic}; LLM-augmented reward and feedback accelerate RL for cyber operations \cite{tholl2025large}; and our prior framework couples a frozen LLM planner with a trainable RL executor, improving multi-stage attacks against the CAGE-4 expert defender \cite{shaikh2026spie}. These works establish that the paradigms can complement one another, yet each is validated in a single environment. What remains missing is a controlled comparison, under one protocol, of the two homogeneous end-to-end architectures, an all-RL and an all-LLM planner-executor agent, across structurally different cyber environments, together with an analysis of where each fails along the attack progression. We aim to fill this gap.

\section{Architectures and Environments}
\subsection{Two Homogeneous Planner-Executor Architectures}
\begin{figure}[t]
\centering
\begin{tikzpicture}[
  font=\footnotesize,
  role/.style={draw=black!55, rounded corners, align=left, inner sep=3.5pt,
               minimum height=6mm, text width=5.2cm, thick},
  arr/.style={-{Latex[length=2mm]}, semithick},
]
\node[role, fill=rlblue!3](plan){\textbf{Planner}\hfill\textit{every $k$ steps}\\[1pt]
  {\scriptsize{\color{rlblue}\textbf{RL+RL:}}~high-level PPO policy}\\
  {\scriptsize{\color{llmorange}\textbf{LLM+LLM:}}~served planner LM}};
\node[role, fill=rlblue!3, below=7mm of plan](exec){\textbf{Executor}\hfill\textit{every step}\\[1pt]
  {\scriptsize{\color{rlblue}\textbf{RL+RL:}}~low-level PPO policy}\\
  {\scriptsize{\color{llmorange}\textbf{LLM+LLM:}}~served executor LM}};
\node[role, fill=black!7, align=center, below=7mm of exec](env){\textbf{Environment}: CAGE-4 / Cyberwheel\\[1pt]{\scriptsize $+$ expert (H-MARL / HRL) blue defender}};
\draw[arr] (plan) -- node[right, fill=yellow!22, rounded corners, inner sep=2pt, font=\scriptsize]{intent $\langle$tactic, zone, risk$\rangle$} (exec);
\draw[arr] (exec) -- node[right, font=\scriptsize]{action} (env);
\draw[arr] (env.west) to[out=180,in=180,looseness=1.2] node[left, font=\scriptsize]{observation} (plan.west);
\end{tikzpicture}
\caption{Both architectures share one two-level pipeline; only the paradigm filling each box changes: {\color{rlblue}blue} for RL+RL (PPO policies) and {\color{llmorange}orange} for LLM+LLM (served language models). The two levels communicate through a compact discrete intent schema $\langle$tactic, zone, risk$\rangle$; the planner is re-invoked every $k$ steps, while the executor acts every step.}
\label{fig:arch}
\end{figure}
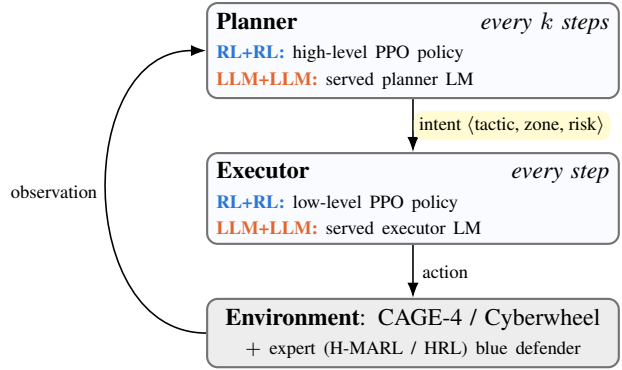

Both agent architectures share the two-level pipeline in Figure~\ref{fig:arch}. A \emph{planner} observes a summary of the network state and emits a strategic subgoal, a compact discrete intent of (tactic, target zone, risk); an \emph{executor} maps that intent and the current state to a primitive action drawn from the environment's action set. The planner is re-invoked every $k$ steps or upon a defender event, while the executor acts every step. What makes each architecture homogeneous is that both levels use the same agent type, either RL or LLM.

In RL+RL, the planner is a high-level RL policy and the executor a low-level RL policy with a multi-discrete action head; a 128-dimensional intent encoder conditions the executor on the planner's subgoal, and both levels are trained with PPO \cite{schulman2017ppo}. A single shared planner and executor are conditioned per operator zone and drive up to six red operator slots. This constitutes a two-level hierarchical agent, which is distinct from a flat single-level RL controller.

\textcolor{black}{The red agent receives no reward of its own; its reward is the negation of the frozen defender's reward, so it gains exactly what the defender loses, and no separate attacker reward is defined. In CAGE-4 that quantity is the defender's per-zone penalty. The defender loses between $1$ and $10$ points whenever the attacker impacts a host or makes a legitimate (green) service or local-work request fail, weighted by zone (operational zones are penalized most) and mission phase. The earlier kill-chain stages earn nothing until they produce an impact. The low-level executor is credited only the penalty from its operator's subnet (zone-credited), and the high-level planner receives the total, discounted over its $k$-step window. Cyberwheel provides no per-zone defender reward, so it specifies the kill-chain reward directly. The executor is rewarded the phase magnitude of each clean action ($+1$ reconnaissance, $+2$ port scan, $+5$ host discovery, $+10$ lateral movement, $+20$ privilege escalation, $+50$ impact), densified by a potential-based progress term ($50\cdot\text{impacted}+10\cdot\text{rooted}+1\cdot\text{accessed}$). The planner is rewarded the same magnitude weighted by the target subnet's criticality and by whether the executor advanced in the subnet it selected ($\times 1$ if so, $\times 0.1$ otherwise), or a $-0.5$ penalty when no progress is made.}

{\color{black}In LLM+LLM, both the planner and executor are language models. The planner reasons over a text summary of the state to choose the subgoal, and the executor translates that subgoal into a concrete primitive action.} The models are frozen, with no task-specific training. Table~\ref{tab:setup} lists the models used in each configuration; one configuration (gpt-oss) uses a larger model for the planner and a smaller model for the executor, while the other two use a single domain-specialized model for both roles.

{\color{black}We evaluate three planner-invocation schedules: \emph{fixed} (replan every $k=10$ steps), \emph{event} (replan early when a defender event fires, otherwise at $k$), and \emph{learned} (a value head decides at each step whether replanning is worthwhile, following the when-to-ask idea~\cite{when2ask}). RL+RL is evaluated under all three schedules, whereas the three LLM+LLM configurations use the fixed schedule, so each setting contributes six configurations (Table~\ref{tab:setup}).}

\begin{table}[t]
\color{black}
\caption{Experimental setup. Each of the six red agent configurations is run in all three settings, giving $3\times6=18$ cells of $n=200$ episodes each (3{,}600 episodes in total). Each RL+RL row is a separate two-level PPO policy trained under its own replanning schedule; Foundation-Sec and DeepHat employ the same model in both roles.}
\label{tab:setup}
\centering
\footnotesize
\setlength{\tabcolsep}{4.5pt}
\begin{tabular}{@{}lccc@{}}
\toprule
\textbf{Setting} & \textbf{Hosts} & \textbf{Steps/episode} & \textbf{Defender} \\
\midrule
CybORG CAGE-4 & multi-subnet & 500$^{\dagger}$ & H-MARL expert \\
Cyberwheel-100 & 100 & 100 & frozen HRL \\
Cyberwheel-1010 & 1010 & 150 & frozen HRL \\
\bottomrule
\end{tabular}

\smallskip
\begin{tabular}{@{}lll@{}}
\toprule
\textbf{Configuration} & \textbf{Planner $\to$ Executor} & \textbf{Replanning} \\
\midrule
RL+RL (fixed) & RL policy $\to$ RL policy & every $k{=}10$ steps \\
RL+RL (event) & RL policy $\to$ RL policy & on a defender event \\
RL+RL (learned) & RL policy $\to$ RL policy & learned gate \\
gpt-oss & gpt-oss-120b $\to$ gpt-oss-20b & every $k{=}10$ steps \\
Foundation-Sec & Foundation-Sec-8B-Reasoning & every $k{=}10$ steps \\
DeepHat & DeepHat-V1-7B & every $k{=}10$ steps \\
\bottomrule
\end{tabular}

{\par\smallskip\scriptsize\raggedright
$^{\dagger}$CAGE-4 episodes run at most 500 steps and may end earlier; both Cyberwheel horizons are fixed.
\par}
\end{table}

\subsection{Environments and Defenders}
CybORG CAGE-4 is a compact, multi-subnet enterprise network partitioned into multiple security zones (including restricted and operational zones), under partial observability over 500-step episodes~\cite{baillie2020cyborg,kiely2025marl}. It provides a dense, kill-chain-aligned attacker signal and is defended by an H-MARL expert policy, the strongest known published CAGE-4 defender~\cite{singh2025hierarchical}. {\color{black}Cyberwheel~\cite{cyberwheel} is evaluated on 100 and 1010 hosts against frozen hierarchical RL defenders, with each episode run for the same number of steps the corresponding defender was trained on: 100 steps for 100 hosts and 150 steps for 1010 hosts.} The two scales differ structurally: the 1010-host topology has no cross-subnet interfaces, so lateral movement out of the entry subnet is impossible, whereas the 100-host topology permits it. Together the environments span a compact, densely rewarded network and a large, sparsely connected one.

\subsection{Evaluation Protocol and Metrics}
We unify the notion of success across architectures with a single predicate: an episode succeeds if at least one host is \emph{disrupted}, i.e., an Impact or Degrade action lands. We report the success rate over $n=200$ episodes per configuration. Total reward is excluded from every cross-architecture comparison because it is shaped differently per architecture. To localize failures we report the \emph{kill-chain reach}, the fraction of episodes reaching each stage (discover, scan, compromise, root, impact), and the planner cost in calls per episode. Every reported number was recomputed from the raw per-episode logs by an independent code path.

\section{Cross-Environment Results}
\begin{figure}[t]
\centering
\includegraphics[width=0.65\linewidth]{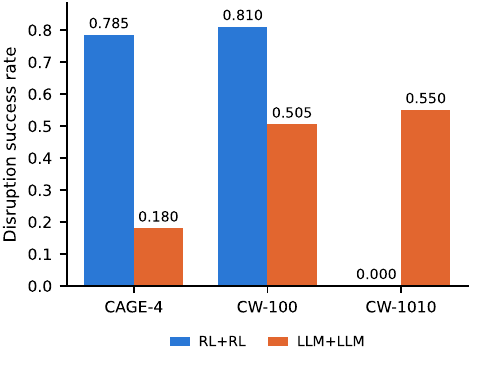}
\caption{\color{black} The winning architecture inverts with the environment: the best RL+RL schedule versus the best LLM+LLM model per setting: CAGE-4 and Cyberwheel (CW).}
\label{fig:inversion}
\end{figure}

Tables~\ref{tab:cc4} and~\ref{tab:cw} and Figure~\ref{fig:inversion} report disruption success for all 18 configurations, and the main result is an environment-dependent inversion. On CAGE-4, RL+RL wins decisively: success rises monotonically across schedules, 68.5\%, 72.0\%, and 78.5\% (fixed, event, learned), against 18.0\% for the strongest language model. {\color{black}On Cyberwheel, the outcome differs across the two network scales: at 100 hosts RL+RL also wins, reaching 81.0\% (fixed) against 50.5\% for Foundation-Sec, whereas at 1010 hosts the ordering reverses, with Foundation-Sec reaching 55.0\% and every RL schedule 0.0\%. Each architecture collapses on the other's home environment, so neither is uniformly stronger; the two are complementary.}

\begin{table*}[t]
\caption{CybORG CAGE-4 results (200 episodes per configuration). Kill-chain columns report the percentage of episodes reaching each stage, with Disruption corresponding to success rate. Degrade reports episodes achieving degradation, while OT-Impact, Zones, and Detect report successful impacts on operational technology, disrupted zones, and blue detections per episode. Best disruption is shown in bold.}
\label{tab:cc4}
\centering
\footnotesize
\setlength{\tabcolsep}{2.5pt}
\begin{tabular}{@{}lrrrrrrrrrr@{}}
\toprule
& \multicolumn{7}{c}{\textbf{Kill-chain reach (\%)}} & \multicolumn{3}{c}{\textbf{CAGE-4 mechanics (mean/episode)}} \\
\cmidrule(lr){2-8}\cmidrule(l){9-11}
\textbf{Configuration} & \textbf{Discover} & \textbf{Scan} & \textbf{Compromise} & \textbf{Root} & \textbf{Impact} & \textbf{Disruption} & \textbf{Degrade} & \textbf{OT-Impact} & \textbf{Zones} & \textbf{Detect} \\
\midrule
RL+RL (fixed) & 100.0 & 100.0 & 100.0 & 76.5 & 65.0 & 68.5 & 61.5 & 2.27 & 0.95 & 0.11 \\
RL+RL (event) & 100.0 & 100.0 & 100.0 & 81.0 & 67.5 & 72.0 & 66.0 & 2.31 & 1.06 & 0.08 \\
RL+RL (learned) & 100.0 & 100.0 & 100.0 & 81.5 & 71.0 & \textbf{78.5} & 70.5 & 2.34 & 1.16 & 0.10 \\
\midrule
gpt-oss & 100.0 & 100.0 & 100.0 & 100.0 & 2.0 & 2.0 & 0.5 & 0.02 & 0.02 & 0.28 \\
Foundation-Sec & 100.0 & 100.0 & 100.0 & 96.0 & 15.0 & 18.0 & 9.0 & 0.70 & 0.21 & 0.15 \\
DeepHat & 100.0 & 100.0 & 100.0 & 20.0 & 0.0 & 0.0 & 0.0 & 0.00 & 0.00 & 0.09 \\
\bottomrule
\end{tabular}
\end{table*}

\begin{table}[t]
\caption{Cyberwheel results (200 episodes per configuration). Kill-chain columns report the percentage of episodes reaching each stage. Note that Cyberwheel has no Degrade action. CAGE-4-specific metrics are omitted. Best disruption per scale is shown in bold.}

\label{tab:cw}
\centering
\footnotesize
\setlength{\tabcolsep}{1.5pt}
\begin{tabular}{@{}lrrrrrr@{}}
\toprule
\textbf{Configuration} & \textbf{Discover} & \textbf{Scan} & \textbf{Compromise} & \textbf{Root} & \textbf{Impact} & \textbf{Disruption} \\
\midrule
\multicolumn{7}{@{}l}{\textit{100-host}} \\
\quad RL+RL (fixed) & 100.0 & 100.0 & 100.0 & \textcolor{black}{89.0} & \textcolor{black}{81.0} & \textcolor{black}{\textbf{81.0}} \\
\quad RL+RL (event) & 100.0 & 100.0 & 100.0 & \textcolor{black}{65.5} & \textcolor{black}{62.5} & \textcolor{black}{62.5} \\
\quad RL+RL (learned) & 100.0 & 100.0 & 100.0 & \textcolor{black}{60.0} & \textcolor{black}{51.0} & \textcolor{black}{51.0} \\
\addlinespace[1pt]
\quad gpt-oss & 100.0 & 100.0 & 100.0 & 55.0 & 33.5 & 33.5 \\
\quad Foundation-Sec & 100.0 & 100.0 & 100.0 & 85.0 & 50.5 & \textcolor{black}{50.5} \\
\quad DeepHat & 100.0 & 100.0 & 100.0 & 0.0 & 0.0 & 0.0 \\
\midrule
\multicolumn{7}{@{}l}{\textit{1010-host}} \\
\quad RL+RL (fixed) & 100.0 & 100.0 & 100.0 & \textcolor{black}{3.5} & \textcolor{black}{0.0} & \textcolor{black}{0.0} \\
\quad RL+RL (event) & 100.0 & 100.0 & 100.0 & 0.0 & 0.0 & 0.0 \\
\quad RL+RL (learned) & 100.0 & 100.0 & 100.0 & 0.0 & 0.0 & 0.0 \\
\addlinespace[1pt]
\quad gpt-oss & 100.0 & 100.0 & 100.0 & 44.5 & 19.5 & 19.5 \\
\quad Foundation-Sec & 100.0 & 100.0 & 100.0 & 91.0 & 55.0 & \textbf{55.0} \\
\quad DeepHat & 100.0 & 100.0 & 100.0 & 0.0 & 0.0 & 0.0 \\
\bottomrule
\end{tabular}
\end{table}

\textcolor{black}{\textbf{The specific model, not the architecture, determines LLM success.} Foundation-Sec is the strongest language model in all three settings. gpt-oss reaches 33.5\% and 19.5\% disruption on the two Cyberwheel scales but only 2.0\% on CAGE-4, and DeepHat reaches 0.0\% in all three of its cells. DeepHat is also a domain-pretrained security model, yet it never succeeds; the outcome therefore depends on the particular model, not completely on domain pretraining or the architecture by itself.}

\textcolor{black}{\textbf{Replanning {\color{black}schedule} is a secondary factor.} {\color{black}On CAGE-4, RL disruption increases across schedules (learned $>$ event $>$ fixed), whereas on Cyberwheel-100 the ordering reverses (fixed $>$ event $>$ learned: 81.0\%, 62.5\%, and 51.0\%) and on Cyberwheel-1010 every schedule stays at 0.0\%. RL disruption varies across schedules, but no schedule changes which architecture leads in a setting; on Cyberwheel-100 the learned schedule leads Foundation-Sec only narrowly (51.0\% versus 50.5\%).} The CAGE-4 ordering comes from the separately trained executor policies, not from plan quality.}

\section{Discussion: Where Each Architecture Breaks Along the Attack Progression}
\label{sec:analysis}

\textbf{Every configuration reaches compromise; the break is at escalation.} In every one of the 18 configurations, discovery, scanning, and host compromise are reached in 100\% of episodes; no architecture fails early. The divergence appears one stage higher, at privilege escalation, and takes two environment specific forms.

{\color{black}\textbf{RL on Cyberwheel-1010: a privilege-escalation wall.} At 1010 hosts, the RL agent discovers 21 to 50 hosts per episode on average and reliably compromises its entry host, but it escalates to root in at most 3.5\% of episodes and disrupts no host in any episode. Because privilege escalation acts on the already compromised host and requires no lateral movement, this is a failure of the access-to-escalation conversion that the policy does not learn on this network, not of exploration; nor is the network itself a barrier to escalation, since gpt-oss and Foundation-Sec reach root in 44.5\% and 91.0\% of episodes on it. At 100 hosts, by contrast, the RL agent clears this stage, reaching root in 60.0\% to 89.0\% of episodes and disruption in 51.0\% to 81.0\%.}

\textbf{Language models on CAGE-4: a root-to-impact cliff.} The opposite failure appears for the all-LLM agent. gpt-oss reaches root in 100\% of episodes but converts only 2\% to impact; Foundation-Sec roots in 96\% and lifts impact reach to 15\%. These agents obtain a privileged foothold almost everywhere and weaponize it almost nowhere.

\textbf{DeepHat: breadth without depth.} DeepHat compromises 7.81 hosts per episode on CAGE-4, more than any other configuration, yet roots 0.21 and impacts 0.00; on Cyberwheel it flatlines at exactly 1.00 compromised host and never reaches root. It climbs the chain broadly but stalls before disruption in every setting, one stage short on CAGE-4 and two on Cyberwheel, evidence that the escalation-to-impact stages are gated by model capability rather than by architecture.

\begin{figure}[t]
\centering
\includegraphics[width=0.65\linewidth]{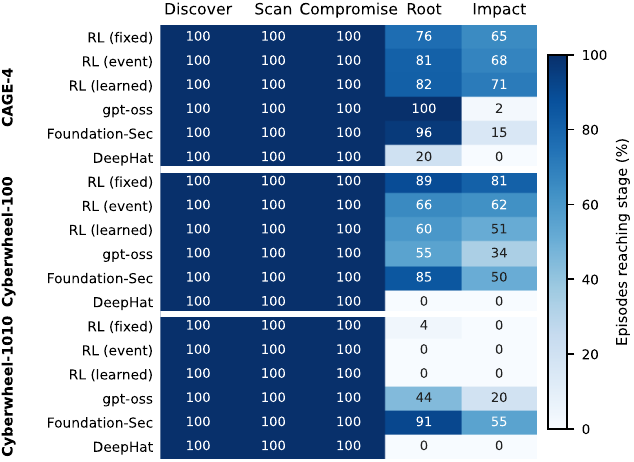}
\caption{Kill-chain reach for all 18 configurations: percent of the 200 episodes reaching each stage (darker $=$ reached). Every configuration reaches compromise in 100\% of episodes (the solid left block) and the divergence is confined to root and impact. \textcolor{black}{RL stalls at escalation on Cyberwheel-1010; gpt-oss roots on CAGE-4 but does not weaponize (100\% root, 2\% impact); RL+RL (fixed) carries Cyberwheel-100 and Foundation-Sec carries Cyberwheel-1010 furthest through impact (gpt-oss also reaches impact on both scales, at 33.5\% and 19.5\%).}}
\label{fig:killchain}
\end{figure}

\begin{figure}[t]
\centering
\includegraphics[width=0.65\linewidth]{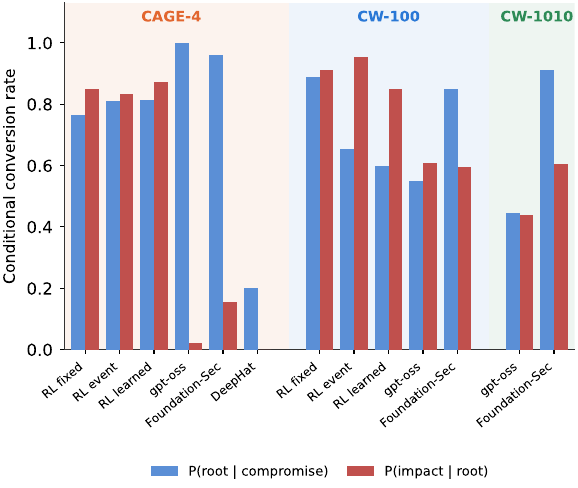}
\caption{Conditional conversion at the two escalation transitions, P(root $\mid$ compromise) and P(impact $\mid$ root), for the \textcolor{black}{13} configurations that reach root in at least 40 of 200 episodes (the rest reach root too rarely to define the conversion; Figure~\ref{fig:killchain}). \textcolor{black}{RL clears both transitions on CAGE-4 and Cyberwheel-100, but on CAGE-4 the language models weaponize root only rarely (gpt-oss 2\%);} the same language models clear the second transition on Cyberwheel. Which transition breaks depends on the environment, not the architecture.}
\label{fig:cascade}
\end{figure}

{\color{black}The two escalation transitions can be read as conditional conversions (Figure~\ref{fig:cascade}), shown for the 13 configurations that reach root in at least 40 of the 200 episodes; the RL Cyberwheel-1010 and DeepHat Cyberwheel cells are absent because they reach root too rarely to define the second conversion, the escalation wall of Figure~\ref{fig:killchain}. The CAGE-4 RL agents clear both transitions (root given compromise up to 81.5\%, impact given root up to 87.1\%), as do the Cyberwheel-100 RL agents (up to 89.0\% and 95.4\%); on CAGE-4, gpt-oss clears the first but falls off the second (100.0\% then 2.0\%),} and Foundation-Sec shows the same CAGE-4 weakness more mildly (96.0\% then 15.6\%); yet the same two language models clear the second transition on Cyberwheel (impact given root 0.44--0.61 across the two scales). Where an architecture does convert root to impact, the final disruptive step is near-free. In Cyberwheel, the all-LLM agent's impact-reach equals its disruption-reach at each scale (50.5\% at 100 hosts, 55.0\% at 1010). The environment, not the architecture, decides which transition breaks.

\begin{figure}[t]
\centering
\includegraphics[width=0.65\linewidth]{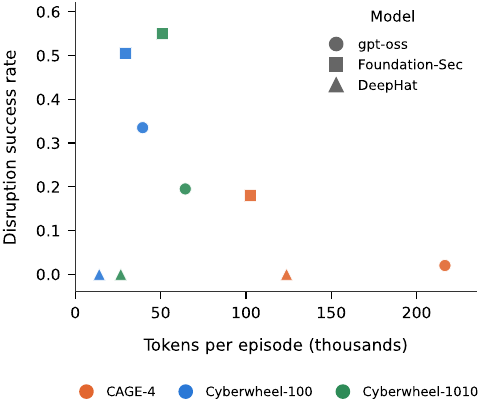}
\caption{Each point is one configuration, where color marks the setting and shape the model. The CAGE-4 language models are the most expensive, 102k-217k tokens per episode, yet disrupt at most 18\% (gpt-oss and DeepHat, the priciest, at most 2\%), \textcolor{black}{whereas Foundation-Sec leads the language models on both Cyberwheel scales at 29k-51k tokens.} Capability and environment, not token budget, decide the outcome. Token counts partly reflect horizon: CAGE-4 episodes run 500 steps versus 100-150 for Cyberwheel.}
\label{fig:tokens}
\end{figure}

\textbf{Efficiency.} Figure~\ref{fig:tokens} shows the all-LLM agent's token cost against disruption: the most token-hungry configurations achieve the least, and RL+RL spends no tokens at all.

\textbf{Is the plan load-bearing?} A plan corruption probe on the CAGE-4 RL agent shows that scrambling the strategic subgoal leaves the executor's action counts byte-identical. \textcolor{black}{In this environment, the planner is \emph{not} load-bearing, and the CAGE-4 schedule ordering of the preceding section reflects the separately trained executor policies rather than plan quality.} This sharpens the case for characterizing the homogeneous architectures end to end, since it identifies precisely which level a hybrid would need to improve.

\textbf{Implications for defense.} The shared failure stage is directly actionable. Because both architectures reach host compromise in almost every episode and diverge along the escalation-to-impact transition, detection and hardening concentrated there, rather than on reconnaissance or initial access, yield the most resilience against autonomous red agents of either paradigm. The environment properties that decide the inversion also predict the threat. \textcolor{black}{A compact, densely instrumented network, or a mid-scale one such as the 100-host Cyberwheel network, is most exposed to a learned agent that has practiced the escalation chain, whereas a large enterprise such as the 1010-host Cyberwheel network is most exposed to a language model agent that supplies escalation tactics from prior knowledge.}

\textbf{Limitations.} Three limits bound our claims. First, we study two simulators across three settings, so our claim rests on the escalation mechanism rather than on breadth of coverage. Second, reward functions were empirically tuned, but alternative reward formulations and sensitivity to reward terms were not systematically explored. Third, rewards are architecture-specific and therefore excluded from cross-architecture performance rankings.

\section{Conclusion and Future Work}
{\color{black}We compared hierarchical red team architectures, RL+RL and LLM+LLM, across CybORG CAGE-4 and Cyberwheel under one disruption metric and found that neither dominates. The stronger architecture inverts with the environment. An all-RL agent wins the compact, densely rewarded CAGE-4 and the 100-host Cyberwheel network, and a cyber-pretrained all-LLM agent wins the large, escalation-gated 1010-host Cyberwheel network. Decomposing the attack progression revealed that every architecture reaches host compromise, and privilege escalation is the stage at which each fails, i.e., RL does not learn to escalate on the 1010-host Cyberwheel network, and the language models gain root but rarely convert it to impact on CAGE-4. The practical reading is a conditions-based guideline, as follows: (1) prefer an RL learned policy when the reward is dense and escalation is reachable from the reward signal, and (2) prefer a capable, cyber-pretrained language model when the network is large and escalation depends on prior tactical knowledge the reward never supplies.}

These results answer the question that motivated the study. \textcolor{black}{A conclusion drawn in a single environment does not reliably transfer,} and the two paradigms are complementary rather than uniformly ordered. This complementarity makes a hybrid architecture attractive, and our analysis identifies where to act (the escalation stage) and which level to strengthen. Building on prior work coupling an LLM planner with an RL executor in one environment~\cite{shaikh2026spie}, we aim to evaluate this hybrid paradigm across environments and explore value-guided decoding that reranks subgoals with a learned critic to adapt planner and executor choices in response to observed performance shortfalls in the deployed environment.

\section*{ACKNOWLEDGMENTS}
This work was supported in part by the U.S. Military Academy (W911NF-25-2-0008), the Defense Advanced Research Projects Agency (HR0011-24-2-0004), and the U.S. Army Research Office (W911NF-24-1-0038). The views and conclusions expressed in this paper are those of the authors and do not reflect the official policy or position of the U.S. Military Academy, U.S. Army, Army Research Office, U.S. Department of War, or U.S. Government. The U.S. Government is authorized to reproduce and distribute reprints for Government purposes notwithstanding any copyright notation herein.

\end{document}